\documentstyle[preprint,aps,tighten,epsf,floats]{revtex}
\newcommand{\ben}{\begin{displaymath}}
\newcommand{\een}{\end{displaymath}}
\newcommand{\be}{\begin{equation}}
\newcommand{\ee}{\end{equation}}
\newcommand{\bea}{\begin{eqnarray}}
\newcommand{\eea}{\end{eqnarray}}

\begin{document}
\draft
\title{On a description of Gravitational Interaction between moving objects in terms of Spatial Gravitational Fields}
\author{ W.D.Flanders\footnote{Department of Epidemiology and Biostatistics, Emory University, Atlanta, GA 30322}, G.S.Japaridze\footnote{Center for Theoretical Studies of Physical Systems, Clark Atlanta University, 223 James P. Brawley Drive, Atlanta, GA 30314; japar@ctsps.cau.edu}}
\maketitle

\begin{abstract}
Analyzing two simple experimental situations we show that from Newton's law 
of gravitation and Special Relativity it follows that the motion of particle in
an external gravitational field can be described in terms of effective spatial fields
which satisfy Maxwell-like system of equations and propagate with the speed of
light.  The description is adequate in a linear approximation in gravitational field and in a
first order in $v^{2}/c^{2}$.
\end{abstract}
\newpage 
\section{Introduction}
In this paper we discuss how the gravitational interaction for an objects with non-zero
velocities can be  described in terms of effective spatial  fields.
Namely, we will show that the  force acting on a particle which moves in
an external gravitational field is given by  the expression
\begin{equation}
\vec F=m\vec g+m \vec v\times\vec {\cal B} ,
\label{lorentz}
\end{equation}
where $\vec g$ is the gravitational field accounting for  Newton's gravitational
law for a particle at rest (see below (\ref{newton})) and the effective field 
${\cal B}$, appearing due to the Special Relativity effects  like the magnetic field in 
electrodynamics, satisfies the relation (below we assume that $\vec g$ -field is time independent)
\begin{equation}
curl\;\vec {\cal B}=-\eta\vec \phi,\; i.e.\;  \oint \vec {\cal B}d\vec l=-\eta \int \vec \phi \vec dA
\label{B}
\end{equation}
In (\ref{B}) $\vec \phi$ is the flow of the unit of mass per unit of time and 
$dl$ and $dA$ stand for the line and area elements.

We will show that
\begin{equation}
\eta=\frac{4\pi\;G_{N}}{c^{2}},
\label{eta}
\end{equation}
where $G_{N}\approx 6,66\cdot 10^{-11}\;Nm^{2}kg^{-2}$ is a Newton's constant and
$c$ is the speed of light.
The small value
of $4\pi\;G_{N}/c^{2}$ explains why, in contrast to their electromagnetic
counterparts, "gravimagnetic" effects  caused by ${\cal B}$  are weak for moderate
values of masses and velocities.

We consider the case of a weak gravitational field and neglect higher powers of
$v^{2}/c^{2}$. In terms of General Relativity we would say that
the curvature of space time nearly vanishes, so that Special Relativity can be
applied with accuracy $1-g_{00}\simeq 1$, $g_{00}$ being the $00$ component of the metric tensor
$g_{\mu\nu}$.
When $g_{00}\ll 1$ the curvature is almost zero, but nearly vanishing deviation
from the flatness of space-time still leads to noticeable acceleration,
described with accuracy $v^{2}/c^{2}$ by Newton's gravitational law  (Dirac 1976, Landau 1962).

Below in section $II$ we consider two {\it gedanken} 
experiments with the point particle and mass flow. We will evaluate 
$\eta$ comparing the result of the first experiment with the expression (\ref{lorentz}) 
and then show that the results for the experiment $2$ are described by (\ref{lorentz})
with the value of $\eta$ obtained from the analysis of experiment $1$. We 
derive equations for the fields $\vec g$ and $\vec {\cal B}$ which 
are similar to Maxwell's equations for the electromagnetic field.

In section $III$ we summarize our results and discuss the limitations of the suggested approach.

The concept of spatial gravitational forces modelled after the electromagnetic Lorentz 
force has a long history and many names associated with it (Holzmuller 1870, Tisserand 1872,  Zel'manov 1956, Cataneo 1958, Bel 1959, Jantzen 1990, Damour 1991,
Bonnor 1995, Braginsky 1997, Mashoon 1997, Dunsby 1997, Maartens 1997) . In this paper we  consider spatial gravitational fields in the most elementary way
and show that even in such  a simplified scheme gravitational phenomena can be analyzed with  the 
accuracy $o(v^{2}/c^{2})$ without invoking equations of  General Relativity. 

\section{experiments with point particle and mass flow and the field equations}
\subsection{Description of Experiments}
In this section we consider two {\it gedanken} experiments: 1) point particle moving
between two infinite pipes which carry a mass flow, and, 2) one pipe, moving towards 
the particle. We analyze these two experiments using only Special Relativity and
Newton's gravitational law
\begin{equation}
\vec F_{12}=G_{N}\frac{m_{1}m_{2}}{r^{3}_{12}}\vec r_{12}
\label{newton}
\end{equation}

The set up for  experiment $1$ is two infinite, straight, massless 
pipes on a plane and a  point particle between them.
Each of the pipes is parallel to the $y$ axis, 
crossing the $x$ axis at $x=\pm b$, and  each  
carries a fluid which flows with velocity $v_{f}$ relative to the pipe. The fluid 
in the left pipe 
moves in the positive $y$ direction,
the fluid in the right pipe  moves in the negative $y$ direction.
The linear density (we neglect the pipe cross-section) of  fluid at rest is 
$\sigma$. The particle with mass $m$ moves along $y$ axis with the velocity
$v_{p}=v_{f}$.

Let us calculate the net force $F$ acting on the particle in the frame of 
reference where particle is (momentarily) at rest - reference frame comoving with the particle.
In this frame of reference the particle  lies  between two continuous mass 
flows with the linear densities $\sigma$ and $\sigma\gamma_{V}$ where $\gamma_{V}$ is a
Lorentz factor accounting the relativistic length contraction
\begin{equation}
\gamma_{V}=(1-\frac{V^{2}}{c^{2}})^{-1/2}
\label{gamma}
\end{equation}
In (\ref{gamma}) $V=-\frac{2v_{f}}{1+v^{2}_{f}/c^{2}}$ is the velocity of a fluid from a right pipe in 
this frame of reference.
The forces exerted from the pipes have only an $x$ components and the straightforward 
calculation leads to the following expression for the force acting on a particle in the frame 
where particle is at rest:
\begin{equation}
F=\frac{2G_{N}m\sigma}{b}(\gamma_{V}-1)\approx \frac{4G_{N}m\sigma}{b}\frac{v^{2}_{f}}{c^{2}}
\label{f}
\end{equation}
To obtain (\ref{f}) we use expression (\ref{newton}), e.g. the force exerted from the left flow 
(second term of (\ref{f})) is 
\begin{equation}
F_{L}=-\int^{+\infty}_{-\infty}dy\frac{G_{N}m\sigma b}{(b^{2}+y^{2})^{3/2}}=
-2G_{N}\frac{m\sigma}{b}
\label{dana}
\end{equation}
The magnitude of the force exerted from the right flow is given by (\ref{dana}) using substitution 
$\sigma\rightarrow \sigma\gamma_{V}$ 
(since we consider infinite pipes
there are no any  boundary effects caused by the endpoints of fluid)
and the final result is the expression (\ref{f}).

When the particle is at rest relative to the flow the net force  is zero, as it 
follows from
(\ref{f}).  
Below we will show 
that ${\cal B}\sim v_{f}$ (see (\ref{approxb})) and that expression 
(\ref{f}) is reproduced 
by the $v$-dependent term of the equation (\ref{lorentz}). 

In the second experiment, the particle with coordinates $(x,y,z)$ is at rest 
and 
the pipe oriented 
along $y$ axis moves towards the particle with the velocity $\vec
v=(-v_{p},0,0)$. The fluid  with 
density $\rho$ moves with the velocity $\vec v_{f}=(0,-v_{f},0)$ relative to
the pipe.

We omit lengthy but straightforward calculations and report only the results for 
$\vec a$ - the acceleration 
of the particle:
\begin{eqnarray}
a_{x}\equiv \frac{dv_{x}}{dt}\approx -\frac{2G_{N}\sigma x}{(x^{2}+z^{2})}(1+
\frac{v^{2}_{f}}{2c^{2}})\cr\cr a_{y}\equiv \frac{dv_{y}}{dt}\approx
\frac{2G_{N}\sigma v_{p}v_{f}x}{c^{2}(x^{2}+z^{2})}\cr\cr a_{z}\equiv
\frac{dv_{z}}{dt}\approx -\frac{2G_{N}\sigma z}{(x^{2}+z^{2})}(1+
\frac{v^{2}_{p}}{2c^{2}}+\frac{v^{2}_{f}}{2c^{2}}), 
\label{a}
\end{eqnarray}
where $x$ is  $\gamma_{v_{p}}$ times the distance from the pipe to the particle 
in the reference  frame where particle is at rest.
Expressions (\ref{a}) are obtained by integration similar to 
that (\ref{dana}) which is based on 
(\ref{newton}), taking 
into account the length contraction of a small element of fluid with mass $\rho dy$ and 
neglecting higher orders of $v^{2}/c^{2}$.

\subsection{Evaluation of $\eta$ from the results of  Experiment $1$}  

To obtain the value of $\eta$, appearing in the relation (\ref{B}) it is necessary 
to consider the problem in a reference frame where particle has a non-zero velocity. 
The simplest solution is  provided by the original frame of reference described previously when
particle and the fluid from a left pipe move along $y$ axis with velocities $v_{p}$ and $v_{f}$ correspondingly, 
$v_{p}=v_{f}$,  
and the fluid from the right pipe moves with the velocity $-v_{f}$.

From the symmetry arguments  it follows that the net field $\vec g$ is zero (particle is 
between two sources with the same linear densities $\sigma\gamma$), so only  the  second term of (\ref{lorentz}) contributes. 
According to (\ref{lorentz}) the force acting on the  particle 
is directed along $x$ axis  and its magnitude is 
$mv_{p} {\cal B}$. This expression already assumes that 
$\vec {\cal B}\vec v_{f}=\vec {\cal B}\vec v_{p}=0$, i.e.  $\vec {\cal B}$ 
is perpendicular to the pipes just as the magnetic field given by Ampere's circuital law 
is perpendicular to the current (Landau 1962).  
In experiment $1$ we have ${\partial \vec g}/{\partial t}=0$ (pipes are at rest), so  
we can use (\ref{B}) to calculate the value of $\vec {\cal B}$. 
Particle is at rest relative to the fluid in the left pipe, so 
it "feels"  field $\cal {\cal B}$ generated only by the pipe from the right.
 Using the integral 
relation  (\ref{B}) and for the relative velocity
$V=-\frac{2v_{f}}{1+v^{2}_{f}/c^{2}}$ we obtain
\begin{equation}
2\pi b {\cal B}=-\eta\sigma\gamma V=\eta \sigma \frac{2v_{f}}{1+\frac{v^{2}_{f}}{c^{2}}} 
 \approx 2\eta\sigma v_{f},
\label{gia}
\end{equation} 
i.e.
\begin{equation}
{\cal B}\approx \eta\sigma v_{f}/\pi b
\label{approxb}
\end{equation}
and the force acting on a moving particle in the original reference frame will be (taking into account $v_{p}=v_{f}$):
\begin{equation}
{\cal F}=mv_{p}{\cal B}=\frac{\eta m \sigma v^{2}_{f}}{\pi b}
\label{origforce}
\end{equation}
In order to establish the value of $\eta $ we have to compare $F$ and ${\cal  F}$ - force calculated in a two 
reference frames.
Reference frames move relative each other along $y$ axis with velocity $V$ and the force is directed along 
$x$ axis. Thus we need Lorentz transformations in its  vectorial form:
\begin{equation}
t^{\prime}=\gamma\biggl ( t-\frac{(\vec V \vec r)}{c^{2}} \biggr);\;
\vec r^{\prime}=\gamma\biggl ( \gamma^{-1}\vec r-\vec Vt+
(1-\gamma^{-1})\frac{(\vec V\vec r)\vec V}{V^{2}}  \biggr)
\label{genlorentz}
\end{equation} 
From (\ref{genlorentz}) and the expression for the force
\begin{equation}
\vec F=\frac{d\vec p}{dt}=\frac{d(m\gamma \vec v)}{dt}=m\gamma \vec a+
m\gamma^{3}\frac{(\vec v \vec a)\vec v}{c^{2}},
\label{relatforce}
\end{equation}
it follows that up to order $v^{4}/c^{4}$ (let us remind that $F$ itself is of order $v^{2}/c^{2}$) we have 
$F={\cal F}$. Therefore, equating expressions for $F$ and ${\cal F}$ we obtain

\begin{equation}
\eta=\frac{4\pi\;G_{N}}{c^{2}}\biggl ( 1+o(\frac{v^{2}}{c^{2}}) \biggr )
\label{eta1}
\end{equation}

\subsection{description of Experiment $2$ in terns of $\vec g$ and $\vec {\cal B}$}
After the value of $\eta$ is established 
we demonstrate that the results for experiment $2$ are described by 
(\ref{lorentz}). To do so, instead of straightforward but 
lengthy
arguments we assume that when ${\partial {\vec g}}/{\partial t}\neq 0$ 
(in the reference frame used t{o describe  
experiment $2$, the  pipe has non-zero velocity, so ${\vec g}$ is now time dependent) 
expression (\ref{B}) is modified as
\begin{equation}
curl \vec {\cal B}=\frac{1}{c^{2}}\frac{\partial {\vec g}}{\partial {t}}-
\eta \vec \phi,\;i.e.\;\oint \vec {\cal B}d\vec l=\int (\frac{1}{c^{2}}\frac{\partial {\vec g}}{\partial {t}}-\eta \vec \phi)d\vec A
\label{1}
\end{equation}
We will verify that (\ref{lorentz}) and (\ref{1}) lead to the same expressions for the 
force as did calculation using only Newton's law and Special Relativity.

The components of $\vec {\cal B}$ can be calculated from (\ref{1}). 
For $\vec g$ we use $\vec g=\vec a$  with $\vec a$ given by (\ref{a}) 
where $d\vec g/dt=(d\vec g/dx)\cdot (dx/dt)$, and 
assume that 
fields vanish at infinity. The choice $\vec g=\vec a$ is justified since in the original 
frame of reference particle is at rest and its (instantaneous) acceleration is 
defined by a "$\vec g$  term "of (\ref{lorentz}). Using $\eta=4\pi G_{N}/c^{2}$,  in a leading 
approximation in $v^{2}/c^{2}$ we obtain:
\begin{eqnarray}
{\cal B}_{x}\approx \frac{2G_{N}\sigma v_{f}z}{c^{2}(x^{2}+z^{2})}\cr\cr
{\cal B}_{y}\approx -\frac{2G_{N}\sigma v_{p}z}{c^{2}(x^{2}+z^{2})}\cr\cr
{\cal B}_{z}\approx -\frac{2G_{N}\sigma v_{f}x}{c^{2}(x^{2}+z^{2})}
\label{b}
\end{eqnarray}

Now we are in a position to check that experiment $2$ can be described by  
$\vec F=m\vec g+m \vec v\times\vec {\cal B}$ .

When the particle is at rest in the reference frame used to describe experiment $2$,
$\vec F=m\vec g$, which is trivially consistent with the equations  
(\ref{a}), since we have defined $\vec g=\vec a$.

To account the effect caused by the "${\cal B}$ - term" of (\ref{lorentz}), we consider the case when  particle in the original frame of 
reference moves  with the non-zero velocity 
$\vec u=(u,0,0)$ . The $y$ component of (\ref{lorentz}) is:
$F_{y}=mg_{y}-m{\cal B}_{z}u$. Direct substitution for ${\cal B}_{z}$ (and
$g_{y}=a_{y}$) results in \begin{equation}
F_{y}=m\frac{2G_{N}\sigma v_{p}v_{f}x}{c^{2}(x^{2}+z^{2})}+
mu\frac{2G_{N}\sigma v_{f}x}{c^{2}(x^{2}+z^{2})}
\label{f2y}
\end{equation}
To compare (\ref{f2y}) with the expression calculated in the framework of Newtonian approximation, 
$F_{y}=ma_{y}$, we 
need the value of $a_{y}$. From (\ref{a}), acceleration in case of 
$\vec u=0$, it follows that when a particle has non-zero $\vec u=(u,0,0)$, the value of 
$a_{y}$ can be obtained by substitution of $v_{p}+u$ for $v_{p}$ in (\ref{a}):
$v_{p}\rightarrow v_{p}+u$ which leads to
\begin{equation}
F_{y}=ma_{y}(v_{p}\rightarrow v_{p}+u)=m\frac{2G_{N}\sigma (v_{p}+u)v_{f}x}{c^{2}(x^{2}+z^{2})}
\label{f22y}
\end{equation}
As it is clear, (\ref{f2y}) and (\ref{f22y}) agree. 

Next we consider the $z$ component. Using (\ref{b}) for $B_{y}$ we obtain:
\begin{equation}
F_{z}=mg_{z}+muB_{y}
=-m\frac{2G_{N}\sigma z}{x^{2}+z^{2}}(1+\frac{v^{2}_{p}}{2c^{2}}+
\frac{v^{2}_{f}}{2c^{2}})-
m\frac{2G_{N}\sigma v_{p}z}{c^{2}(x^{2}+z^{2})}u
\label{f2z}
\end{equation}
Since the particle moves with the velocity $\vec u=(u,0,0)$ we have 
$a_{z}=F_{z}/\gamma_{u}m$ (see (\ref{relatforce})): 
\begin{equation}
a_{z}\approx -\frac{2G_{N}\sigma z}{x^{2}+z^{2}}(1+\frac{v^{2}_{p}}{2c^{2}}+
\frac{v^{2}_{f}}{2c^{2}}-\frac{u^{2}}{2c^{2}})-
\frac{2G_{N}\sigma v_{p}z}{c^{2}(x^{2}+z^{2})}u,
\label{a2z}
\end{equation}  
where $\gamma^{-1}_{u}\equiv \sqrt {1-u^{2}/c^{2}}\approx 1-u^{2}/2c^{2}$.  

Now we have to compare this expression 
with the one 
for $a_{z}$ from (\ref{a}), 
calculated from Newton's law and Special Relativity  - acceleration of a particle
in the reference frame comoving with the particle . We replace $v_{p}\rightarrow v_{p}+u$
in the expression (\ref{a}) for $a_{z}$ to obtain:
\begin{equation}
a_{z}(v_{p}+u)=-\frac{2G_{N}\sigma
z}{x^{2}+z^{2}}(1+\frac{v^{2}_{p}}{2c^{2}}+\frac{v^{2}_{f}}{2c^{2}}+
\frac{v_{p}u}{c^{2}}+\frac{u^{2}}{2c^{2}}) \label{f22z}
\end{equation}
Expression (\ref{f22z}) gives the acceleration of the particle in a reference frame 
moving along the $x$ axes with velocity $v_{p}+u$ relative to the pipe. On the other 
hand, expression (\ref{a2z}) corresponds  to the acceleration in the reference frame
moving along the $x$ axes with velocity $v_{p}$ relative to the pipe. To compare 
(\ref{a2z}) and (\ref{f22z}), we use (\ref{genlorentz})  to transform  acceleration from the 
reference frame used in (\ref{f22z}) to that used in (\ref{a2z}): 
$a_{z}\rightarrow a^{\prime}_{z}\gamma^{-2}_{u}$. 
This transformation introduces the term 
$-u^{2}/c^{2}$ so that the two expressions for acceleration now agree.

A similar analysis for the case when the velocity of the particle is along the $y$ axis, 
$\vec u=(0,u,0)$ again confirms that the expression $\vec F=m\vec g+m\vec v\times \vec {\cal B}$ 
can be used  to describe the motion of a particle in a gravitational field.

\subsection{Equations for $\vec g$ and $\vec {\cal B}$}
Besides equation (\ref{1}) which was postulated (and subsequently 
verified to describe experiment 
$2$ in a self-consistent way) it is possible work out two more relations for the fields 
$\vec {\cal B}$ and $\vec g$. 

First of all, from the expressions  (\ref{b}) it follows that 
${\partial {{\cal B}_{x}}}/{\partial x}+{\partial {{\cal B}_{y}}}/{\partial y}+
{\partial {{\cal B}_{z}}}/{\partial z}=0$, i.e.
\begin{equation}
div\;\vec {\cal B}=0
\label{divb}
\end{equation}

Next we compare $curl\;{\vec g}$ and ${\partial {\vec {\cal B}}}/{\partial {t}}$. For the $y$ 
component we obtain:
\begin{equation}
\frac{{\partial {B_{y}}}}{{\partial {t}}}=\frac{{\partial }}{{\partial {t}}}
\Biggl( -\frac{2G_{N}\rho v_{p}z}{
c^{2}(x^{2}+z^{2})} \Biggr)=
\frac{2G_{N}\rho v^{2}_{p} 2zx}{
c^{2}(x^{2}+z^{2})^{2}}
\label{amper1y}
\end{equation}
Straightforward calculation of the $y$
component of $curl\;{\vec g}$ (as before, we take $\vec g=\vec a$, for $\vec a$ see 
(\ref{a})) results in
\begin{eqnarray}
\frac{{\partial {g_{x}}}}{\partial z}-\frac{{\partial {g_{z}}}}{\partial x}=
\frac{{\partial }}{\partial z}\Biggl( -\frac{2G_{N}\rho x}{
(x^{2}+z^{2})}
(1+\frac{v^{2}_{f}}{2c^{2}})
 \Biggr)-\frac{{\partial }}{\partial x}
\Biggl( -\frac{2G_{N}\rho z}{
(x^{2}+z^{2})} \Biggr)\approx 
-\frac{2G_{N}\rho 2zx}{
(x^{2}+z^{2})^{2}}\frac{v^{2}_{p}}{c^{2}},
\label{amper2y}
\end{eqnarray} 
i.e. $(curl\;{\vec g})_{y}=-{\partial {\cal B}_{y}}/{\partial {t}}$.

Consideration of the $x$ and $z$ components show that the relation 
\begin{equation}
curl\;{\vec g}=-\frac{{\partial {\vec {\cal B}}}}{{\partial {t}}}
\label{amper}
\end{equation}
is  valid. Also, from the definition of $\vec g$ we have $div\;{\vec g}=4\pi G_{N}\rho$ where $\rho$ is a 
regular three dimensional density.

Summarizing, the equations for $\vec g$ and $\vec {\cal B}$ are as follows:
\begin{eqnarray}
div\;{\vec g}=4\pi G_{N}\rho,\; curl\;{\vec g}=-{\partial {\vec {\cal B}}}/{\partial {t}}
\cr\cr
div\;\vec {\cal B}=0,\; curl\;\vec {\cal B}=\frac{1}{c^{2}}\frac{\partial {\vec g}}{\partial {t}}-\frac{4\pi\;G_{N}}{c^{2}}\vec j
\label{equations}
\end{eqnarray}
where $\rho$ is the mass density and $\vec j$ is the mass density flow.
In case of
experiments $1$ and $2$ $\vec j=\vec \phi=\rho \vec v_{f}$.

It is now straightforward to obtain 
the wave equations for the case $\rho=0$, $\vec j=0$:
\begin{equation}
\frac{1}{c^{2}}\frac{{\partial^{2} {\vec{g}}}}{{\partial t^{2}}}=\bigtriangledown^{2} {\vec g},\;
\frac{1}{c^{2}}\frac{{\partial^{2} {\vec {\cal B}}}}{{\partial t^{2}}}=
\bigtriangledown^{2} \vec {\cal B},
\label{wave}
\end{equation}
i.e. free waves propagate with speed of light.

\section{discussion}
We have demonstrated that the gravitational 
force acting on a point particle with mass $m$ and velocity $\vec v$ is given by the 
expression 
\begin{equation}
\vec F=m\vec g+m\vec v\times \vec {\cal B}
\label{force}
\end{equation}
with $ \vec g$ and $ \vec {\cal B}$ satisfying the system of equations similar to  the Maxwell 
equations. 

The approximation we used is that gravitational field is weak enough so that space-time 
is approximately euclidean and the velocities are small enough so that 
higher powers of $v^{2}/c^{2}$ are negligible. In the framework of this approximation 
the force obtained from Newton's law (\ref{newton}) and the Special relativity is described 
by (\ref{force}), i.e. motion of particle is given by an expression  similar to 
the Lorentz force for a charged particle in an external electromagnetic field. The similarity 
is caused by neglecting the effects of self-interaction for gravitational field, 
corresponding to a non linearity of Einstein's equations. 
In case of classical electromagnetism the linear approximation to field equations is 
well justified in a sense that phenomena with characteristic action substantially exceeding $\hbar$, 
$\hbar$ being the Planck's constant, are described by Maxwell's and Lorentz's equations (Landau 1962)
and in electromagnetic phenomena quantum effects manifest themselves  earlier than 
effects caused by  a non linear corrections to Maxwell's equations.
Intuitively it becomes clear when one compares
 electron's Compton wave length $r_{q}=\hbar /mc$ and its classical electromagnetic radius
$r_{e}=e^2/mc^{2}$: from the value of the ratio $r_{q}/r_{e}=\hbar c/e^{2}\approx 137$ it follows 
that the quantum effects, namely the pair production occurs at a distance
which is 137 times more than the distance at which the classical field singularities become 
relevant and  when it becomes necessary to modify classical theory, e.g. to  introduce non linear terms in field equations.

Theory of gravity provides us with an opposite feature -  "classical radius" $r_{g}=2G_{N}m/c^{2}$ 
appearing in the Schwarzschild's  metric (Dirac 1976, Landau 1962) greatly exceeds Compton  
wavelength - $r_{g}/r_{q}=2m^{2}/M^{2}_{Pl}$, where the Planck mass $M_{Pl}\approx
10^{-5}gr$.  Therefore in describing the motion of bodies with $m\gg M_{Pl}$ it is vital 
to consider the exact classical equations of motion (Einstein's non linear equations)  - 
quantum effects are negligible at this scale.
The self-interaction plays a decisive role 
in describing basic phenomena of light deflection or precession of perihelion of 
planetary motion (Dirac 1976, Landau 1962). This features of a motion  in  a {\it static}  gravitational field
can not be described by (\ref{force}) - the "${\cal \vec B}$ term" is absent for a static source.
Therefore, fields $\vec g$ and $\vec {\cal B}$, describing gravitational interaction of a moving 
objects in a linear approximation,  
can be treated only as an effective fields and the limitation of our approach 
manifests itself in a degrees of freedom: $6$ components of $\vec g$ and $\vec {\cal B}$ 
of course are not enough to describe the degrees of freedom of a gravitational field.

Despite that the relevancy of the linear approximation is questionable, 
approximation  (\ref{force}) can be still useful  
for describing interaction of moving bodies :
from the General Relativity it follows  that
the exact expression  for the force exerted on  point particle moving in an external stationary 
field is given by expression similar to (\ref{force}) (Landau 1962):
\begin{equation}
\vec F=-mc^{2}\;\vec \bigtriangledown ln\;\sqrt{-g_{00}}+
mc\sqrt{-g_{00}}\vec v\times curl\;\vec {\cal G},
\label{landau}
\end{equation}
where ${\cal G}_{\alpha}\equiv -g_{\alpha 0}/g_{00}$, $\alpha$ stands for a spatial part of metric  and
$g_{\mu\nu}$ is a metric tensor. When $g_{00}=-1-2\Phi/c^{2}$, where 
$\Phi$ is a scalar potential, $\Phi/c^{2}\ll 1$,  the first term of the r.h.s. of (\ref{landau}) is the 
same as the first term of the r.h.s. of 
(\ref{force}). To reproduce 
the second 
term of (\ref{force}) which  includes the  field $\vec {\cal B}$, that is to express 
$\vec {\cal B}$ in terms of $g_{\mu\nu}$ it would be  necessary to solve 
Einstein's equations.  
At the moment, we 
know of no 
solutions for the Einstein's equations for experiments $1$ and $2$, but based on  our phenomenological
consideration  we believe that the equation similar to  (\ref{force}) can be derived
from the equations of General Relativity.

Let us note that though fields $\vec g$ and $\vec {\cal B}$ satisfy wave equations (\ref{wave}), they do not transform as an antisymmetric tensor of rank $2$, i.e. they do not transform as the electromagnetic field $F_{\mu\nu}\sim (\vec E,\;\vec H)$.  If one attempts to postulate that the {\it exact} expression for the force acting on a test particle is given by  (\ref{force}) or (\ref{landau}) then it turns out that in order to maintain expression (\ref{force}) fields $\vec g$ and $\vec {\cal B}$ transform like non-tensor quantities (Dirac 1976). This is the price one has to pay when attempting to describe gravitational interaction in terms of $6$ degrees of freedom. The non-tensor feature of transformation is most transparent from (\ref{landau}): identifying $\vec g$ with the first term of r.h.s. of (\ref{landau}) it follows that at $x^{\mu}\rightarrow x^{\mu}+\xi^{\mu}(x)$ in the expression for the transformed $\vec g$ there arises extra term
\begin{equation}
\delta g_{i}(x)=\frac{\partial}{\partial x^{i}}\Biggl ( \frac{g^{\mu0}}{g^{00}}\frac{\partial \xi^{\mu}}{\partial x^{\mu}}  \Biggr)ln\;\sqrt{-g_{00}},
\label{extra}
\end{equation}
which can not be compensated by the transformation of a "$\vec {\cal B}$-term" of (\ref{landau}). Therefore eq. (\ref{force}) can not hold in any reference frame, for {\it any} velocities. In Einstein's equations extra terms similar to (\ref{extra}) are compensated by coordinate transformations of General Relativity and as a result, equations of gravitational field and the requirement of general covariance form a self consistent mathematical scheme (Dirac 1976, Landau 1962).

In approximation used in this paper (linearized equations and lowest order in $v^2/c^2$) fields $\vec g$ and $\vec {\cal B}$ transform as $\vec E$ and $\vec H$. This statement is true only in lowest order in $v^2/c^2$. Straightforward calculation shows that (\ref{a}) and (\ref{b}) transform exactly as $\vec E$ and $\vec H$ transform in the lowest order in $v^2/c^2$. Therefore, in the framework of approximation used the description based on (\ref{force}) and (\ref{equations}) is self consistent.

As we already have mentioned, $\vec g$ and $\vec {\cal B}$ are effective fields,
even from the point of view of classical theory. Nevertheless,
equation (\ref{force})  can be applied to a rather wide class 
of phenomena in  problem of describing the motion in external gravitational field
after the fields $\vec g$ and $\vec {\cal B}$ are known. The advantage of using (\ref{force}) and 
(\ref{equations})   is in  
their simplicity in comparison with the problem of solving equations of General Relativity.

\section{Acknowledgment}
One of us (G.S.J) is grateful to D. Finkelstein and 
G. Mezincescu for useful discussions.

\newpage
\begin{center}
REFERENCES
\end{center}
1.  L.Bel, {\it C.R. Acad. Sci.}, {\bf 247} 1094 (1959).\\
2.  W.Bonnor, {\it Class. Quantum Grav.} {\bf 12}, 499 (1995).\\
3.  V.B.Braginsky, C.M.Caves, K.S.Thorne, {\it Phys.Rev} {\bf 15}, 2047 (1977).\\
4.  C.Cataneo, {\it Nuovo Cimento}, {\bf 11} 733 (1958). \\
5.  T.Damour, M.Soffel, C.Xu, {\it Phys.Rev.D} {\bf 43}, 3273 (1991).\\
6.  P.A.M.Dirac, {\it General Theory of Relativity}, NY, Willey-Intersience Publication,
{\bf 1976}.\\
7.  P.K.Dunsby, B.A.C.Basset, G.F.R.Ellis, {\it Class. Quantum Grav.} {\bf 14}, 1215 (1997).\\
8.  G.Holzmuller, {\it Z. Math. Phys.} {\bf 15}, 69 (1870).\\
9.  R.T.Jantzen, P.Carini, D.Bini, {\it Ann. Phys.} {\bf 215}, 1 (1990).\\
10. L.D.Landau, E.M.Lifshitz, {\it The Classical Theory of Fields}, Oxford, Pergamon Press, 
{\bf 1962}.\\
11. R.Maartens, G.F.R.Ellis, S.T.C.Siklos, {\it Class. Quantum Grav.} {\bf 14}, 1927 (1997).\\
12. B.Mashoon, J.McClune, H.Quevedo, {\it Phys. Lett.A}, 47 (1997)..\\
13. F.Tisserand, {\it Compt. Rend.}, {\bf 110} 760 (1872).\\
14. A.L.Zel'manov, {\it Sov. Phys. Doklady} {\bf 1}, 227 (1956).

\end{document}